\documentstyle[psfig]{mn}
\def\be{\begin{equation}}

\def\ee{\end{equation}}

\def\etal{et al.}

\def\hMpc{h^{-1}{\rm Mpc}}
\def\hcMpc{h^{-3}{\rm Mpc}^3}
\def\hcMpcinv{h^{3}{\rm Mpc}^{-3}}

\def\kms{\mbox{km s$^{-1}$}}

\def\ten#1{\times 10^{#1}}

\def\refer#1{#1}
\def\refeq#1{\relax}		

\newcommand{\Halpha}{\mbox{H$\alpha$}}
\newcommand{\Hi}{\mbox{H\,{\sc i}}}
\newcommand{\Hii}{\mbox{H\,{\sc ii}}}
\newcommand{\Nii}{\mbox{[N\,{\sc ii}]}}
\newcommand{\Oii}{\mbox{[O\,{\sc ii}]}}
\newcommand{\ew}{\mbox{EW}}
\newcommand{\ewh}{\mbox{EW\,(H$\alpha$)}}
\newcommand{\ewo}{\mbox{EW\,([O\,{\sc ii}])}}

\title[LF and clustering by spectral type]
{Spectral Analysis of the Stromlo-APM Survey II.\\ 
Galaxy luminosity function and clustering by spectral type}

\author[J. Loveday, L. Tresse \& S. Maddox]
{J.~Loveday$^{1}$, 
L.~Tresse$^{2}$ and 
S.~Maddox$^{3,4}$\\
$^1$Department of Astronomy and Astrophysics, University of Chicago, 
5640 S. Ellis Ave, Chicago, IL 60637, USA\\
$^2$Istituto di Radioastronomia del CNR, Via P. Gobetti, 101, 
40129 Bologna, Italia\\
$^3$Institute of Astronomy, Madingley Road, Cambridge, CB3 0HA, UK\\
$^4$School of Physics and Astronomy, University of Nottingham, Nottingham, 
NG7 2RD, UK\\
loveday@oddjob.uchicago.edu, tresse@ira.bo.cnr.it, sjm@ast.cam.ac.uk
}
\begin{document}

\maketitle

\begin{abstract}
We study the luminosity function and clustering properties of subsamples
of local galaxies selected from the Stromlo-APM survey by the rest-frame
equivalent widths of their \Halpha\ and \Oii\ emission lines.
The $b_J$ luminosity function of star-forming galaxies has a significantly 
steeper faint-end slope than that for quiescent galaxies:
the majority of sub-$L^*$ galaxies are currently undergoing significant
star formation.
Emission line galaxies are less strongly clustered, both amongst themselves,
and with the general galaxy population, than quiescent galaxies.
Thus as well as being less luminous, star-forming galaxies also inhabit
lower-density regions of the Universe than quiescent galaxies.
\end{abstract}

\begin{keywords}
cosmology: observations
--- galaxies: clustering
--- galaxies: luminosity function, mass function
--- surveys
\end{keywords}

\section{Introduction}

Important clues to the physics of galaxy formation and evolution may be
obtained by studying the global properties, such as the luminosity function 
and correlation function, of quiescent versus star-forming galaxies.
The most reliable tracer of the formation rate of massive, hot stars is
the flux of the \Halpha\ emission line, directly related to the stellar 
UV ($< 912$ \AA) photoionizing flux (\refer{Kennicutt 1983}).
This line is frequently redshifted out of the observed spectral window,
and so most deep galaxy surveys have instead used the \Oii\ 3727\AA\ line
as a measure of star-formation rate (\refer{Kennicutt 1992}).

The luminosity function of galaxies subdivided by the presence or absence of
the \Oii\ emission line has been calculated in the local Universe
for the Las Campanas Redshift Survey (LCRS, \refer{Lin \etal\ 1996a}) 
and for the ESO Slice Project (ESP, \refer{Zucca \etal\ 1997}).
In both surveys it was found that the faint-end of the galaxy luminosity 
function is dominated by \Oii\ emitters, in other words that presently
star-forming galaxies tend to be less luminous than quiescent galaxies
in both the $b_J$ (ESP) and Gunn-$r$ (LCRS) bands.
These results from \Oii-selected samples are consistent with the recent 
luminosity function estimates from local samples of galaxies selected by 
morphological (eg. \refer{Marzke \etal\ 1998}) and spectral 
(eg. \refer{Bromley \etal\ 1998}, \refer{Folkes \etal\ 1999}) type:
early-type (elliptical and lenticular) galaxies tend to be luminous, 
and late-type (spiral and irregular) galaxies faint.

It is by now also well known (eg. \refer{Davis \& Geller 1976}, 
\refer{Giovanelli \etal\ 1986}, \refer{Iovino \etal\ 1993}, 
\refer{Loveday \etal\ 1995}) that galaxies of early morphological type
cluster together on small scales more strongly than late-type galaxies.
Since emission-line galaxies (ELGs) tend to be of late Hubble type, 
we would expect
ELGs to be more weakly clustered than non-ELGs, and indeed this has been
observed by numerous authors (eg. \refer{Iovino \etal\ 1988},
\refer{Salzer 1989}, \refer{Rosenberg, Salzer \& Moody 1994} and
\refer{Lin \etal\ 1996b}).

In this paper we study the luminosity function and clustering for subsamples
of the Stromlo-APM survey (\refer{Loveday \etal\ 1996}) selected by \Halpha\
and \Oii\ emission-line equivalent widths.
The Stromlo-APM survey is ideal for quantifying the statistical
properties of emission-line versus quiescent galaxies in the local universe 
since it contains a representative
sample of different galaxy types and covers a large volume
$V \approx 1.38 \ten{6}\ \hcMpc$.
Since the red wavelength coverage of Stromlo-APM spectra extends from 
6300--7600 \AA\, we are able to detect the \Halpha\ (6562.82\AA) line, 
when present,
to a redshift $z \la 0.16$, i.e. beyond the maximum distance reached by 
the survey.
Thus for the first time we are able to classify a large, representative 
sample of galaxies by the primary tracer of massive star formation, viz.
the equivalent width of the \Halpha\ emission line.
Measurement of the spectral properties of Stromlo-APM galaxies is discussed
by \refer{Tresse \etal\ (1999)}, hereafter referred to as Paper~1.
The subsamples selected by their emission-line properties are described in
\S\ref{sec:samples}.
The luminosity functions of the different samples are compared in 
\S\ref{sec:LF} and in \S\ref{sec:clust} we present clustering measurements.
We summarize our results in \S\ref{sec:concs}.
Throughout, we assume a Hubble constant of $H_0 = 100 h$ \kms Mpc$^{-1}$ with 
$h=1$ and a deceleration parameter $q_0 = 0.5$.  
The exact cosmology assumed has little effect at redshifts $z \la 0.15$.

\section{Galaxy Samples} \label{sec:samples}

Our sample of galaxies is taken from the Stromlo-APM redshift
survey which covers 4300 sq-deg of the south galactic cap and consists
of 1797 galaxies brighter than $b_J = 17.15$ mag.  The galaxies all
have redshifts $z < 0.145$, and the mean is $\langle z \rangle =
0.051$.  A detailed description of the spectroscopic observations and
the redshift catalog is published by \refer{Loveday \etal\ (1996)}.

Of the 1797 galaxies originally published in the redshift survey, 
82 have $b_J < 15$.
These bright galaxies are excluded from our analysis since they tend to be
saturated on the Schmidt plates and hence have unreliable magnitudes.
Of the remaining 1715 galaxies, 26 have a redshift taken from the literature, 
and for 7 we could not retrieve the spectra because they were
not observed with the Dual-Beam Spectrograph (DBS) of the ANU 2.3-m
telescope at Siding Spring.  
Also excluded were 6 blueshifted spectra, 3 with $c z < 1000$ km s$^{-1}$,
and 2 with too low signal-to-noise.  

The remaining 1671 spectra were flux-calibrated and had their spectral 
properties measured as described in Paper~1.
Flux calibration of our spectra is accurate to $\sim 10$--$20\%$, and so in
the present paper we have restricted our analysis to galaxy samples selected
by the equivalent widths (EWs) of their \Halpha\ and \Oii\ emission lines,
which are insensitive to flux calibration errors.
Note that since the resolution of our spectra has FWHM = 5\AA, the \Halpha\ 
line can always be deblended from the \Nii\ doublet.

\begin{figure}
\psfig{figure=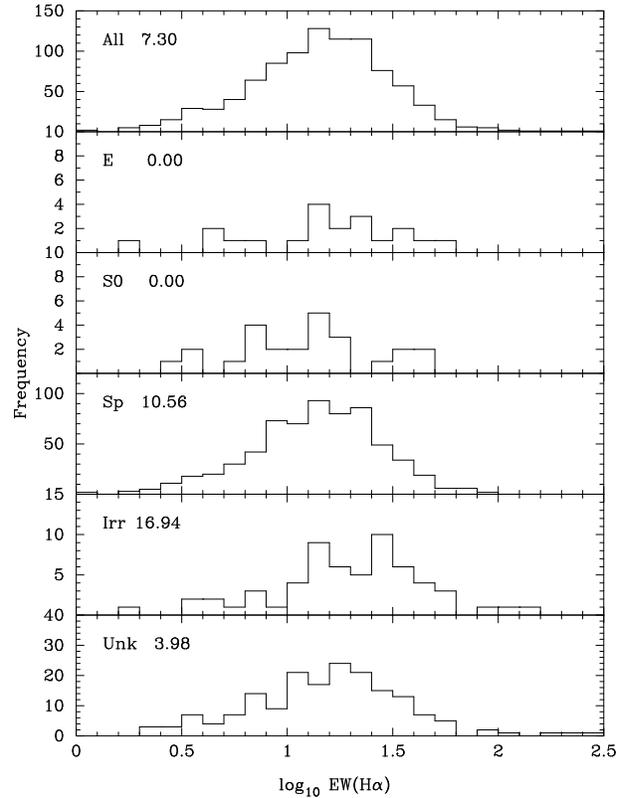,width=90mm}
\caption{Histogram of log \ewh\ for all galaxies and for morphologically
selected subsamples as labeled.  Note that galaxies with no detected \Halpha\
are not shown here.
The median \ewh, including non-detections, is given for each sample after 
the morphological type.
\label{fig:ha_hist}}
\end{figure}

\begin{figure}
\psfig{figure=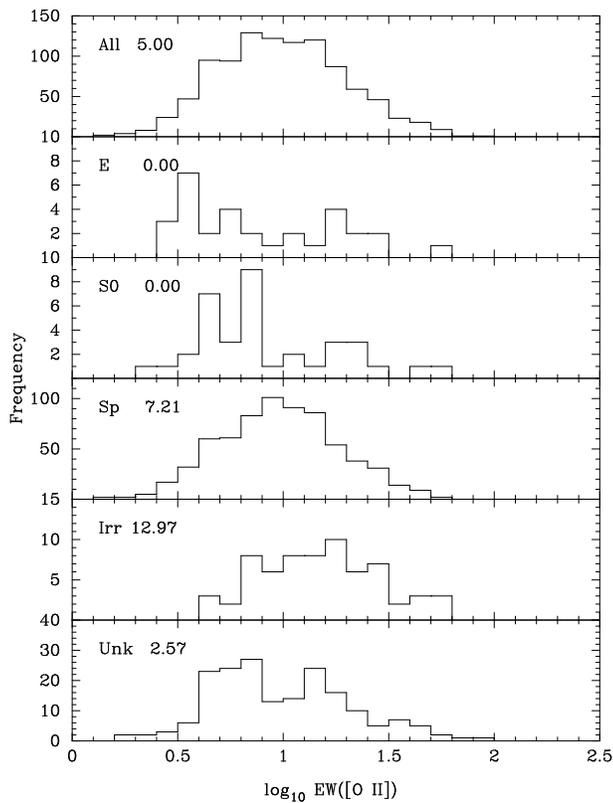,width=90mm}
\caption{Histogram of log \ewo\ for all galaxies and for morphologically
selected subsamples as labeled.  Note that galaxies with no detected \Oii\
are not shown here.
The median \ewo, including non-detections, is given for each sample after 
the morphological type.
\label{fig:oii_hist}}
\end{figure}

Of the 1671 measured galaxies, 11 were not part of our core statistical sample,
either because they had an uncertain redshift or happened to lie in a part of
the sky masked by ``holes'' around bright stars, etc.
Of the remaining 1660 galaxies, 82 could not have \ewh\ measured as 
their redshift places the \Halpha\ line in a small gap in the red part of the 
spectrum from 7000--7020\AA\ (\refer{Loveday \etal\ 1996}).
For an additional 57 spectra, \Halpha\ was seen in emission but could not be
measured due to contamination by a sky line, or some other problem with the
spectrum; \Oii\ lines could not be measured for similar reasons for 5 spectra.
Note that lack of EW measurement, while correlated with redshift, is 
uncorrelated with galaxy morphology, and so we can reliably correct for missing
EW measurements.
We are thus left with a sample of 1521 galaxies which could be analysed
by \ewh, and 1655 which could be analysed by \ewo.
Histograms of log \ewh\ and log \ewo\ are plotted in 
Figures~\ref{fig:ha_hist} and \ref{fig:oii_hist} respectively.

We select galaxy subsamples using measured equivalent widths of the \Halpha\
and \Oii\ emission lines.
The \Halpha\ line is the best tracer of massive star formation (Kennicutt 1983)
but we also select samples using the equivalent width of the \Oii\ line, as
this line allows us to compare with other surveys in which \Halpha\ is not 
always within the wavelength range measured.
The \Halpha\ line is detected with \ew\ $\ge 2$\AA\ in 61\% of galaxies.
Of these emission-line galaxies, half have \ewh\ $ > 15$\AA.
Thus we form three subsamples of comparable size by dividing the sample at
\ewh\ of 2\AA\ and 15\AA.
In the case of the \Oii\ line, 60\% of galaxies have \ew\ $\ge 2$\AA, and
of these half have \ewo\ $\ge 9.6$\AA.
The galaxy samples selected by \Halpha\ and \Oii\ equivalent widths are defined
in Table~\ref{tab:samples}.

\begin{table}
 \begin{center}
 \caption{Spectroscopic subsamples and correlation with morphological type.}
 \label{tab:samples}
 \begin{tabular}{llrrrrrr}
 \hline
 \hline
 Sample & \ewh & E & S0 & Sp & Irr & Unk & Total\\
 \hline
 (a) H-low & $ < 2$ \AA & 125 & 108 & 207 & 10 & 149 & 599\\
 (b) H-mid & 2--15 \AA &   8 &  16 & 340 & 18 &  81 & 463\\
 (c) H-high & $ > 15$ \AA &  11 &   9 & 303 & 41 &  95 & 459\\
 \hline
 Sample & \ewo & E & S0 & Sp & Irr & Unk & Total\\
 \hline
 (d) O-low & $ < 2$ \AA & 120 & 112 & 239 &  8 & 177 & 656\\
 (e) O-mid & 2--9.6 \AA &  19 &  24 & 344 & 19 &  97 & 503\\
 (f) O-high & $ > 9.6$ \AA &  12 &  12 & 339 & 47 &  86 & 496\\
  \hline
  \hline
 \end{tabular}
\end{center}
\end{table}

Most galaxies in the Stromlo-APM survey have had a morphological type 
(elliptical, lenticular, spiral or irregular) assigned
by visual inspection of the galaxy image (\refer{Loveday 1996},
\refer{Loveday \etal\ 1996}).
In Table~\ref{tab:samples} we give the numbers of galaxies of each
morphological type in each spectroscopically selected subsample.
In Figures~\ref{fig:ha_hist} and \ref{fig:oii_hist} we also plot 
the distribution of equivalent widths for these morphologically-selected
subsamples.
The sample labeled ``Unk'' consists of galaxies to which no morphological
classification was assigned.
We see that early-type galaxies dominate when \Halpha\
or \Oii\ emission is not detected and are underrepresented when emission lines
are detected.
Conversely, the number of irregular galaxies increases significantly in the 
spectroscopic samples which show strongest star formation.
Strong star formation is known to disrupt the regularity in the shape of a 
galaxy.
In the deeper universe, the apparent increase in number of irregulars is 
also related to strong star formation (\refer{Brinchmann \etal\ 1998}).
Thus as expected we find a good correlation between morphological types 
and emission line equivalent widths.
Since they can be measured objectively, spectroscopic properties of galaxies
are a more reliable discriminator than visually assigned morphological types.
Moreover, a significant fraction of Stromlo-APM galaxies have no morphological
type assigned (the column marked ``Unk'' in Table~\ref{tab:samples}).
The low median \ewh\ and \ewo\ for these unclassified galaxies compared with
the total sample suggests that many are in fact of early morphological type.
The spectral classification described in this section allows these galaxies
to be assigned to their appropriate class in a quantitative way.

\section{The Galaxy Luminosity Function} \label{sec:LF}

We estimate the $b_J$ luminosity function (LF) for each galaxy subsample using
maximum-likelihood, density-independent methods, so that our results are 
unbiased by galaxy clustering.
We use the \refer{Sandage, Tammann \& Yahil (1979)} parametric 
maximum-likelihood estimator to fit a \refer{Schechter (1976)} function,
\be
\phi(L) dL = \phi^* \left(\frac{L}{L^*}\right)^\alpha 
	    \exp\left(\frac{-L}{L^*}\right) dL.
\label{eqn:schec}
\ee
We correct for random errors in our magnitudes by convolving this
luminosity function with a Gaussian with zero mean 
and rms $\sigma_m = 0.30$ (see \refer{Loveday \etal\ 1992}, hereafter L92,
for details).
We also perform a non-parametric fit to each luminosity function using the
stepwise maximum-likelihood estimator of 
\refer{Efstathiou, Ellis \& Peterson (1988)}.
This estimator calculates $\phi(L)$ in a series of evenly-spaced magnitude bins
and provides a reliable error estimate for each bin by inverting the 
information matrix.
$K$-corrections are applied to each galaxy according to its morphological
classification as E/S0: $4.14 z$, Sp: $2.25 z$, Irr: $1.59 z$, Unk: $2.90 z$.

Before calculating the LF for each spectroscopic subsample defined in 
Table~\ref{tab:samples}, we first checked that the galaxies omitted from
this analysis, ie. those galaxies whose \Halpha\ or \Oii\ emission lines
could not be measured, did not bias the LF measurement relative to the full
Stromlo-APM survey.
The LF estimates using all galaxies except the 194 with no \Halpha\
measurement available and all galaxies except the 60 with no \Oii\ measurement
were indeed both consistent with the full sample.

\begin{table*}
\begin{minipage}{12cm}
 \begin{center}
 \caption{Luminosity function parameters.}
 \label{tab:LF}
 \begin{math}
 \begin{array}{lcccccc}
 \hline
 \hline
 {\rm Sample} & \langle V/V_{\rm max} \rangle & \alpha & M^* & 
 \bar{n} & \phi^* & \rho_L\\
 \hline
 \mbox{(a) H-low} & 0.48 \pm 0.01 & -0.75 \pm 0.28 & -19.63 \pm 0.24 & 
	10.1 \pm 2.5 & 4.5 \pm 1.1 & 5.9 \pm 1.4\\
 \mbox{(b) H-mid} & 0.49 \pm 0.01 & -0.72 \pm 0.29 & -19.28 \pm 0.23 & 
	11.0 \pm 2.8 & 5.4 \pm 1.4 & 5.1 \pm 1.4\\
 \mbox{(c) H-high} & 0.54 \pm 0.01 & -1.28 \pm 0.30 & -19.04 \pm 0.26 & 
	48.0 \pm 15.9 & 8.5 \pm 2.8 & 8.8 \pm 2.9\\
 \mbox{(d) O-low} & 0.51 \pm 0.01 & -0.80 \pm 0.29 & -19.51 \pm 0.22 & 
	13.8 \pm 3.2 & 5.8 \pm 1.3 & 6.8 \pm 1.7\\
 \mbox{(e) O-mid} & 0.49 \pm 0.01 & -0.36 \pm 0.34 & -19.16 \pm 0.22 &  
	7.4 \pm 1.8 & 5.8 \pm 1.3 & 4.9 \pm 1.1\\
 \mbox{(f) O-high} & 0.51 \pm 0.01 & -1.49 \pm 0.26 & -19.49 \pm 0.30 & 
	46.0 \pm 15.9 & 3.7 \pm 1.2 & 7.9 \pm 2.8\\
  \hline
  \hline
 \end{array}
 \end{math}
 \medskip

$\alpha$ is the faint-end slope and $M^*$ the characteristic $b_J$
magnitude of the best-fit Schechter function.
$\bar{n}$ is the space density of galaxies in the range $-22 < M < -15$
and $\phi^*$ is the normalisation of the Schechter luminosity function,
both in units of $10^{-3} \hcMpcinv$.
$\rho_L$ is the luminosity density integrated over the same
magnitude range, in units of $10^7 L_{\sun}\hcMpcinv$.
\end{center}
\end{minipage}
\end{table*}

\begin{figure*}
\centerline{\psfig{figure=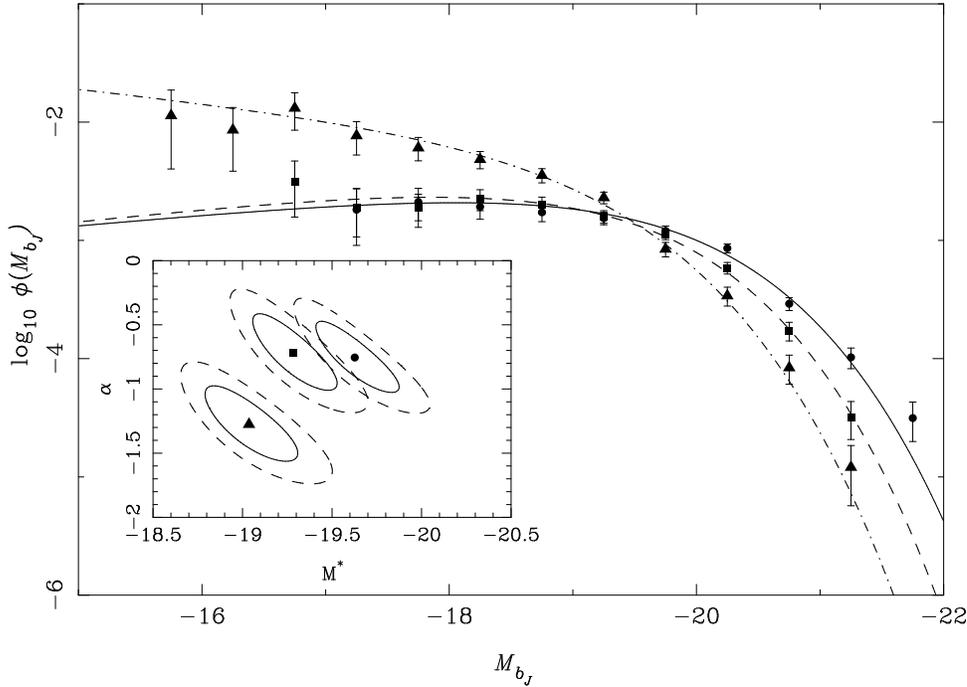,width=140mm,angle=-90}}
\caption{Estimates of the luminosity function for galaxies with no
significant detected \Halpha\ emission (H-low: circles, solid line),
with moderate \Halpha\ emission (H-mid: squares, dashed
line) and with strong \Halpha\ emission (H-high: triangles, dot-dashed line).
The symbols with error bars show the stepwise fit, the curves show the 
Schechter function fits.
For clarity, data points representing fewer than five galaxies have been
omitted from the plot.
The inset shows 1 \& 2$\sigma$ likelihood contours for the best-fit 
Schechter parameters.
\label{fig:lf_ha}}
\end{figure*}

\begin{figure*}
\centerline{\psfig{figure=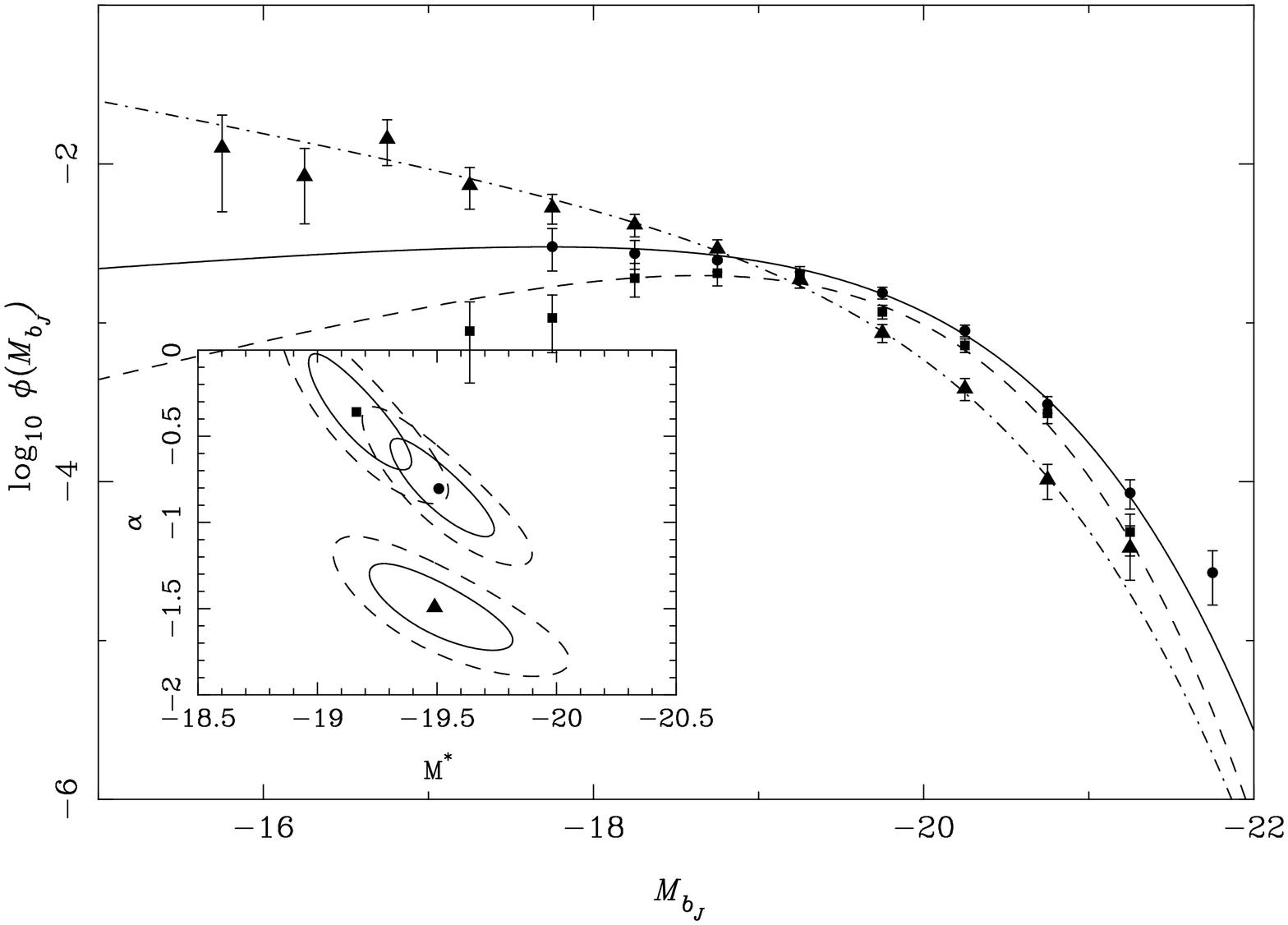,width=140mm,angle=-90}}
\caption{Estimates of the luminosity function for galaxies with no
significant detected \Oii\ emission (O-low: circles, solid line),
with moderate \Oii\ emission (O-mid: squares, dashed
line) and with strong \Oii\ emission (O-high: triangles, dot-dashed line).
The symbols with error bars show the stepwise fit, the curves show the 
Schechter function fits.
For clarity, data points representing fewer than five galaxies have been
omitted from the plot.
The inset shows 1 \& 2$\sigma$ likelihood contours for the best-fit 
Schechter parameters.
\label{fig:lf_oii}}
\end{figure*}

Our estimates of the luminosity function for the \ewh\ selected samples 
are shown in Figure~\ref{fig:lf_ha}.
The inset to this Figure shows the likelihood contours for the best-fit 
Schechter parameters $\alpha$ and $M^*$.
The Schechter parameters and their $1\sigma$ errors (from the bounding box
of the $1\sigma$ error contours) are also listed in Table~\ref{tab:LF}.
Note that the estimates of $\alpha$ and $M^*$ are strongly correlated
and so the errors quoted for $\alpha$ and $M^*$ in the Table are conservatively
large.
We see a trend of faintening $M^*$ and steepening $\alpha$ as \ewh\ increases.
There is a significantly greater contrast between the H-high and H-mid samples
than between the H-mid and H-low samples, despite the rather similar 
distribution of morphological types in the H-high and H-mid samples as compared
with the H-low sample.
This suggests that either there is not a simple one-to-one correlation between
optical morphology and \ewh, or that the larger fraction of Irr galaxies in
the H-high sample are contributing to the steep faint-end slope for this 
sample.

Luminosity function estimates of the \ewo\ selected samples and 
errors in the best-fit
Schechter parameters are shown in Figure~\ref{fig:lf_oii}.
The $1\sigma$ error contours for the O-low and O-mid samples overlap and the
O-high sample does not show a fainter $M^*$ than
non-emission line galaxies.
However, the LF for the O-high sample does have a significantly 
steeper faint-end slope than that for galaxies with only weak or moderate
\Oii\ emission.

The fact that we see a systematic dimming of $M^*$ with emission-line EW
for the \Halpha-selected sample but not for the \Oii-selected sample
is probably due to the fact that \ewh\ is a measure of the 
fraction of ionizing photons from OB stars over the flux from the old stellar
population emitted in the rest-frame $R$ band which forms the continuum at 
\Halpha, while \ewo\ is normalised by the flux from relatively young 
stars (mainly type A). 
Thus \ewh\ is more sensitive to the current star formation rate and hence blue
luminosity enhancement than \ewo.

Note that the LF estimate for late-type galaxies presented by L92
does not have such a steep faint-end slope as we find here for strong 
emission-line galaxies.
In L92 we combined galaxies classified as spiral or irregular
as ``late type'', and so not all of them have strong emission lines.
The faint-end slope for early-type galaxies (L92) was much shallower than
that measured here for galaxies with no emission lines.
At least part of this difference is due to a bias in the morphological
type dependent LFs of L92 due to the tendency of unclassified galaxies 
in the Stromlo-APM survey to be of low luminosity (\refer{Marzke \etal\ 1994}, 
\refer{Zucca \etal\ 1994}).
We avoid this bias with the spectroscopically selected samples analysed here.

The normalisation $\phi^*$ of the fitted Schechter functions was estimated
using a minimum variance estimate of the space density $\bar{n}$ of galaxies
in each sample (\refer{Davis \& Huchra 1982}, L92).
We corrected our estimates of $\bar{n}$, $\phi^*$ and luminosity density 
$\rho_L$ to allow for those galaxies excluded from each subsample.
First, all subsamples were scaled by the factor 1715/1660 to 
account for the 55 galaxies with no \ew\ information available.
Second, all \Halpha\ selected subsamples were scaled by 1660/1578
to account for the 82 galaxies whose \Halpha\ line, if present, would have
fallen in the ``red gap'' (\S\ref{sec:samples}).
Samples H-mid \& H-high were scaled by an additional factor 1578/1521 to
allow for the 57 galaxies in which \Halpha\ was seen, but was not able to be 
measured.
Finally, samples O-mid \& O-high were scaled by 1660/1655 to allow 
for the five galaxies in which \Oii\ was seen but not measured.
Our final estimates of $\bar{n}$, $\phi^*$ and $\rho_L$
are given in Table~\ref{tab:LF}.
The uncertainty in mean density due to ``cosmic variance'' (L92 equation 7)
is $\approx 6\%$ for each sample.
However, the errors in these quantities are dominated by the uncertainty
in the shape of the LF, particularly by the value of the estimated 
characteristic magnitude $M^*$.

Using both \Halpha\ and \Oii\ equivalent widths as indicators of star formation
activity, we find that galaxies currently undergoing significant bursts of 
star formation
dominate the faint-end of the luminosity function, whereas more quiescent
galaxies dominate at the bright end.
This is in agreement with the results of Lin \etal\ (1996a) and 
Zucca \etal\ (1997), but in disagreement with \refer{Salzer (1989)}, who
finds no significant difference in the LF shapes of star-forming and
quiescent galaxies.
As pointed out by \refer{Schade \& Ferguson (1994)}, Salzer's sample is
biased against weak-lined ELGs at low-luminosity, and their reanalysis
of his data correcting for this selection effect does find a steep faint-end
slope for the LF of star-forming galaxies.

The characteristic magnitude $M^*$ for the O-high sample
is about 0.5 mag brighter than that for the H-high sample.
This is probably due to a combination of several factors:
1) A large \Oii\ EW can come from a small \Oii\ flux and a very red continuum
(ie. a small star formation rate and an old stellar population).
2) The correlation between estimated values of faint-end slope $\alpha$
and characteristic magnitude $M^*$ means that the steeper $\alpha$ of the
O-high sample will push the estimated $M^*$ to brighter magnitudes.
3) The errors on $M^*$ are large ($\pm 0.3$ mag), 
and so the H-high and O-high $M^*$ estimates 
disagree only at the 1--2 $\sigma$ level.

\section{Galaxy Clustering} \label{sec:clust}

In this section we measure the clustering properties of the galaxy subsamples.
We measure the auto-correlation function of each sample in redshift space,
and the cross-correlation function of each galaxy sample with all galaxy
types in real space.
For both estimates, we first verified that the 194 galaxies missing \ewh\
measurement and the 60 galaxies missing \ewo\ 
did not bias the measured clustering relative to the complete sample.
Those galaxies excluded because \Halpha\ fell in the ``red gap''
lie at redshifts $z \approx 0.06$--0.07.
Nevertheless, omitting these galaxies did not significantly affect the
measured clustering in real or redshift space.

\subsection{Redshift-Space Correlation Function} \label{sec:clustRed}

\begin{figure}
\centerline{\psfig{figure=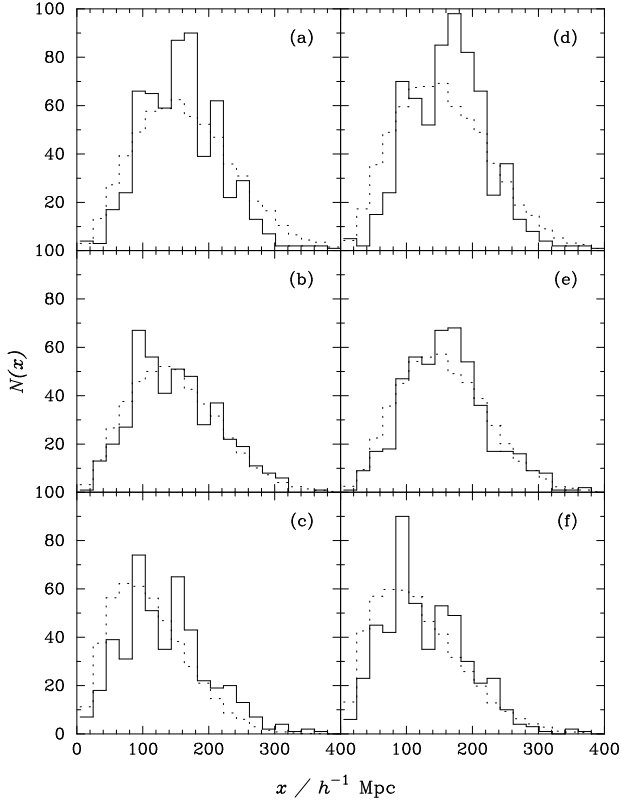,width=90mm}}
\caption{The solid histograms show the observed number-distance $N(x)$ 
distribution for each galaxy subsample defined in Table~\ref{tab:samples}.
The dashed lines show the expected distribution from our luminosity function
fits.
\label{fig:Nofx}}
\end{figure}

We correct for boundary conditions and the survey selection function by
populating the survey volume with a catalogue of $\sim 18,000$ random points 
whose radial density matches that expected for each subsample.
The number-distance distributions for the six galaxy subsamples analysed here
are shown in Figure~\ref{fig:Nofx}.
These plots also show the expected distributions inferred from the
luminosity functions calculated in the previous section.
We see that given the tendency for non-ELGs to be luminous and for ELGs
to be faint, the ELGs are slightly overdense at large distances 
($x \ga 200 \hMpc$) whereas there is an underdensity of non-ELGs at 
similar distances.
This observation is reflected by the increasing $\langle V/V_{\rm max} \rangle$
with \ewh\ seen in Table~\ref{tab:LF}, and is
probably due to evolution in emission line strength
with redshift (eg. \refer{Broadhurst \etal\ 1992}), occurring
at redshifts as low as $z \la 0.15$. 
It is unlikely to be due to the changing projected size of the spectrograph
slit at different redshifts as we demonstrated in Paper~1.
We checked that these discrepancies between observed and expected $N(x)$
distributions did not bias our estimates of $\xi(s)$ by also generating
a random distribution according to
a fourth-order polynomial fit to the observed radial density of each subsample.
Clustering estimates using this random distribution gave results consistent 
with a random distribution generated according to the predicted radial density.

The auto-correlation function of each sample in redshift space is measured
using the estimator
\be
  1 + \xi(s) = \frac{w_{gg}(s) w_{rr}(s)}{[w_{gr}(s)]^2},
  \label{eqn:xi_ni}
\ee
\refer{Hamilton (1993)}.
Here $w_{gg}(s)$, $w_{gr}(s)$ and $w_{rr}(s)$ are the summed products of the 
weights of galaxy-galaxy, galaxy-random and random-random pairs respectively
at separation $s$.
We use the minimum-variance pair weighting given by equation~1 of
\refer{Loveday \etal\ (1995)}, and the reader is referred to that paper
for further details.
Errors are estimated by dividing the survey into four zones of roughly equal
area and calculating the variance in $\xi(s)$ from zone-to-zone.

\begin{figure}
\centerline{\psfig{figure=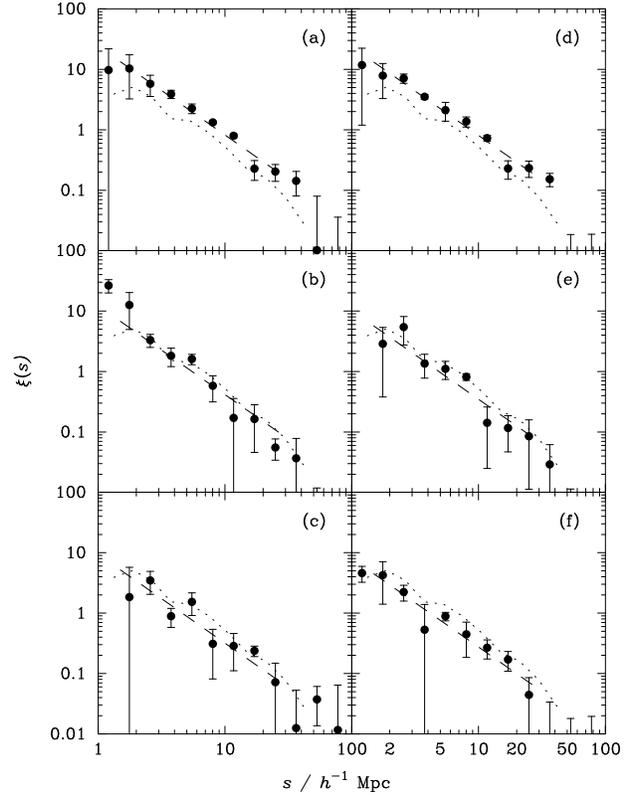,width=90mm}}
\caption{Estimates of the redshift-space correlation function for the
galaxy samples given in Table~\protect{\ref{tab:samples}}.
Error bars show the rms variance from dividing the survey into 4 distinct 
zones.
The dashed line shows the best-fit power-law over the range 1.5--30 $\hMpc$
with the index held fixed at $\gamma_s \equiv 1.47$.
The dotted line shows $\xi(s)$ estimated from the full Stromlo-APM sample
(Loveday \etal\ 1995).
\label{fig:xis}}
\end{figure}

Estimates of $\xi(s)$ are shown in Figure~\ref{fig:xis}.
A power-law $\xi(s) = (s/s_0)^{-\gamma_s}$ was fitted over the range 1.5--30 
$\hMpc$.
For each subsample the estimated power-law slope $\gamma_s$ was formally
consistent with $\gamma_s = 1.47$, measured for the whole Stromlo-APM sample
(Loveday \etal\ 1995).
Since estimates of the index $\gamma_s$ and correlation length $s_0$ are 
strongly correlated, we determined the best fit $s_0$ to each subsample,
keeping the power-law index fixed at $\gamma_s = 1.47$.
The results of these fits are shown by the dashed lines in Figure~\ref{fig:xis}
and the best-fit values of $s_0$ with $1\sigma$ uncertainties (determined
from fitting to each zone separately) are shown in Table~\ref{tab:corr}.

\begin{table}
 \begin{center}
 \caption{Correlation function parameters.}
 \label{tab:corr}
 \begin{math}
 \begin{array}{lccc}
 \hline
 \hline
 \mbox{Sample} & s_0 & \gamma_r & r_0\\
 \hline
 \mbox{(a) H-low} & 8.7 \pm 0.5 & 1.78 \pm 0.08 & 6.0 \pm 1.4\\
 \mbox{(b) H-mid} & 5.5 \pm 0.7 & 1.60 \pm 0.13 & 5.2 \pm 2.0\\
 \mbox{(c) H-high} & 4.6 \pm 0.9 & 1.87 \pm 0.16 & 2.9 \pm 1.9\\
 \mbox{(d) O-low} & 8.6 \pm 1.1 & 1.79 \pm 0.07 & 6.2 \pm 1.8\\
 \mbox{(e) O-mid} & 4.9 \pm 0.6 & 1.64 \pm 0.05 & 4.7 \pm 0.8\\
 \mbox{(f) O-high} & 4.1 \pm 0.9 & 1.78 \pm 0.15 & 2.9 \pm 0.7\\
  \hline
  \hline
 \end{array}
 \end{math}
 \vspace{0.2cm}

$s_0$ is the correlation length measured in redshift space
over the range 1.5--30 $\hMpc$ with the power-law index held fixed at
$\gamma_s \equiv 1.47$.
$\gamma_r$ and $r_0$ are the real-space power-law parameters over
0.2--20 $\hMpc$ determined from cross-correlation with the
2d APM survey (\S\protect{\ref{sec:clustReal}}). 
\end{center}
\end{table}

We see that the correlation length $s_0$ becomes significantly smaller 
in more actively star-forming galaxies, as traced by both \ewh\ and \ewo.
This result is in agreement with the power-spectrum analysis of the
Las Campanas Redshift Survey by \refer{Lin \etal\ (1996b)} who find
that the clustering amplitude of ELGs is only about 70\% that of
the full LCRS sample.
These results are also consistent with those of Rosenberg \etal\ (1994),
Iovino \etal\ (1988) and Salzer (1989), all of whom find that ELGs are less
strongly clustered than quiescent galaxies.
Galaxies with no detected \Halpha\ (H-low) or \Oii\ (O-low) emission have a 
correlation length about twice that of ELG galaxies (H-high and O-high samples).
This is larger than the difference in clustering amplitude determined by 
Lin \etal\ (1996b) from the LCRS, presumably because we have subdivided
galaxies into three EW bins compared to their two EW bins.

\subsection{Real-Space Correlation Function} \label{sec:clustReal}

The estimate of $\xi(s)$ described above is affected by redshift space 
distortions.
On small scales, random, thermal motions tend to decrease galaxy clustering,
whereas on large scales, galaxy streaming motions tend to enhance $\xi(s)$.
In order to avoid the effects of galaxy peculiar velocities, we have calculated
the projected cross-correlation function $\Xi(\sigma)$ of each galaxy subsample
with all galaxies in the APM survey to a magnitude limit of $b_J = 17.15$.
We then invert this projected correlation function to obtain the real space
cross-correlation function $\xi(r)$ of each subsample with the full galaxy 
sample.
This method of estimating $\xi(r)$ is described by 
\refer{Saunders \etal\ (1992)} and by Loveday \etal\ (1995).

\begin{figure}
\centerline{\psfig{figure=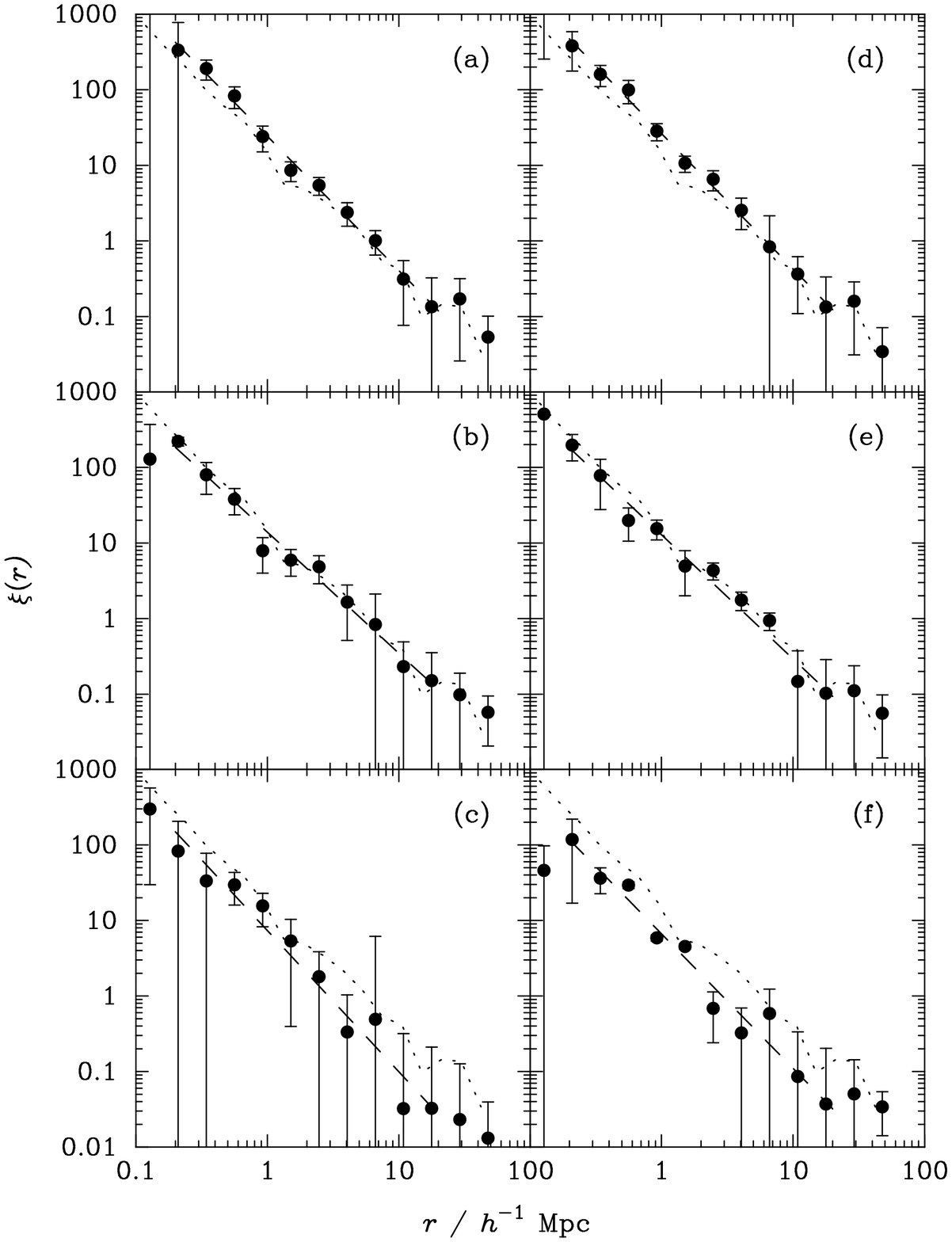,width=90mm}}
\caption{Estimates of the real-space correlation function for the
galaxy samples given in Table~\protect{\ref{tab:samples}}.
Error bars show the rms variance from dividing the survey into 4 distinct 
zones.
The dashed line shows the best-fit power-law over the range 0.2--20 $\hMpc$.
The dotted line shows $\xi(r)$ estimated from the full Stromlo-APM sample
(Loveday \etal\ 1995).
\label{fig:xir}}
\end{figure}

The large number of galaxy pairs used by this estimator allows us to fit a
power-law to the measured cross-correlation function over the range of 
separations 0.2--20 $\hMpc$ and to fit both the power-law index $\gamma_r$
and the correlation length $r_0$.
Our estimates of $\xi(r)$ are plotted in Figure~\ref{fig:xir} and our best-fit
power-laws are tabulated in Table~\ref{tab:corr}.
As in redshift-space, we see that strong emission-line galaxies are more 
weakly clustered than their quiescent counterparts by a factor of about two.

The real space clustering measured for non-ELGs is very close to that measured 
for early-type (E + S0) galaxies, and the clustering of
late-type (Sp + Irr) galaxies lies between that of the
moderate and high EW galaxies (cf. Loveday \etal\ 1995).
Given the strong correlation between morphological type and presence of
emission lines (Table~\ref{tab:samples}) this result is not unexpected.
The power-law slopes are consistent ($\gamma_r = 1.8 \pm 0.1$) 
between the H-low, H-high, O-low and O-high samples.
For the moderate EW galaxies (H-mid and O-mid samples)
we find shallower slopes ($\gamma_r = 1.6 \pm 0.1$).
This is only a marginally significant (1--2 $\sigma$) effect,
but may indicate a deficit of moderately star-forming galaxies principally
in the cores of high density regions, whereas strongly star forming
galaxies appear to more generally avoid overdense regions.

\section{Conclusions} \label{sec:concs}

We have presented the first analysis of the luminosity function and spatial
clustering for representative and well-defined local
samples of galaxies selected by 
\ewh, the most direct tracer of star-formation.
We have also selected galaxies by \ewo, and find broadly consistent results 
between the two tracers of star formation, which is expected from their 
close relation (Kennicutt 1992, Paper~1). 
The observed trend for $M^*$ to fainten systematically with increasing \ewh,
contrasted with the roughly constant $M^*$ with varying \ewo, 
is probably due to \ewh\ being a more reliable indicator of star formation 
rate than \ewo.

Star-forming galaxies are likely to be significantly fainter than their 
quiescent counterparts.
The faint-end ($M \ga M^*$) of the luminosity function is dominated by ELGs
and thus the majority of local dwarf galaxies are currently undergoing 
star formation.

Star-forming galaxies are more weakly clustered, both amongst themselves, 
and with the general galaxy population, than quiescent galaxies.
This weaker clustering is observable on scales from 0.1--10 $\hMpc$.
We thus confirm that star-forming galaxies are preferentially found today in 
low-density environments.

A possible explanation for these observations is that
luminous galaxies in high-density 
regions have already formed all their stars by today, while less luminous 
galaxies in low-density regions are still undergoing star formation.
It is not clear what might be triggering the star formation in these galaxies
today.
While interactions certainly enhance the rate of star formation in some
disk galaxies, interactions with luminous companions can only
account for a small fraction of the total star formation in disk
galaxies today (\refer{Kennicutt \etal\ 1987}).
\refer{Telles \& Maddox (1999)} have investigated the environments of \Hii\
galaxies by cross-correlating a sample of \Hii\
galaxies with APM galaxies as faint as $b_J = 20.5$.
They find no excess of companions with \Hi\ mass $\ga 10^8 M_{\sun}$ near
\Hii\ galaxies, thus arguing that star formation in most \Hii\ galaxies is
unlikely to be induced by even a low-mass companion.

Our results are entirely consistent with the hierarchical picture of
galaxy formation.
In this picture, today's luminous spheroidal galaxies formed from
past mergers of galactic sub-units in high density regions, and produced all
of their stars in a merger induced burst, or series of bursts, over a 
relatively short timescale.
The majority of present-day dwarf, star-forming galaxies in lower density
regions may correspond to unmerged systems formed at lower peaks in the
primordial density field (eg. \refer{Bardeen \etal\ 1986})
and whose star formation is still taking place.
Of course, the full picture of galaxy formation is likely to be significantly
more complicated than this simple sketch, and numerous physical effects
such as depletion of star-forming material and other feedback mechanisms
are likely to play an important role.

\section*{Acknowledgments}

We thank George Efstathiou and
Bruce Peterson for their contributions to the Stromlo-APM survey.

{}

\end{document}